\documentclass[letterpaper, 10 pt, conference]{ieeeconf}  

\IEEEoverridecommandlockouts                              

\usepackage{cite}
\usepackage{amsmath}
\usepackage{amssymb,amsfonts}
\usepackage{graphicx}
\usepackage{textcomp}
\usepackage[dvipsnames]{xcolor}
\usepackage{soul}
\usepackage{tikz}
\usepackage{transparent}        

\usepackage{lipsum}

\usepackage{algorithm}
\usepackage{algpseudocode}

\usepackage{multicol}

\usepackage{pgfplots, pgfplotstable}
\usepgfplotslibrary{groupplots}
\usepgfplotslibrary{fillbetween}
\usetikzlibrary{shapes,arrows,calc,spy}
\pgfplotsset{grid style={dashed}}
\usepackage{subcaption}

\def\BibTeX{{\rm B\kern-.05em{\sc i\kern-.025em b}\kern-.08em
    T\kern-.1667em\lower.7ex\hbox{E}\kern-.125emX}}

\definecolor{camel}{rgb}{0.76, 0.6, 0.42}%

\usepackage[normalem]{ulem}

\usepackage{balance}
\newtheorem{definition}{Definition}[section]

\newcommand{\vect}[1]{\boldsymbol{#1}}
\DeclareSymbolFont{matha}{OML}{txmi}{b}{it}
\DeclareMathSymbol{\varv}{\mathord}{matha}{118}

\makeatletter

\newenvironment{shiftedflalign*}{%
    \start@align\tw@\st@rredtrue\m@ne
    \hskip\parindent
}{%
    \endalign
}
\makeatother

\begin{document}

\title{\LARGE Towards Safe Reinforcement Learning with Reduced Conservativeness: \\ A Case Study on Drone Flight Control}

\author{Loizos Hadjiloizou$^{1}$, Michael C. Welle$^{1}$, Hang Yin$^{2}$, Danica Kragic$^{1}$
\thanks{$^{1}$L. Hadjiloizou, M. C. Welle, and D. Kragic are with the Division of RPL, EECS, KTH Royal Institute of Technology, Lindstedtsvägen 24, 11428 Stockholm, Sweden, {\tt\small \{loizosh, mwelle, danik\}@kth.se}.}        
\thanks{$^{2}$H. Yin is with the Department of Computer Science, University of Copenhagen, Universitetsparken 1, 2100, Copenhagen, Denmark,  {\tt\small hayi@di.ku.dk}}}

\maketitle

\begin{abstract}

Incorporating formal methods into reinforcement learning (RL) has the potential to result in the best of both worlds, combining the robustness of formal guarantees with the adaptability and learning capabilities of RL, though careful design is needed to balance safety and exploration. In this work, we propose a framework to mitigate this loss of exploration while still allowing for the safety of the system to be ensured. Specifically, we introduce a less restrictive method that can reduce the conservativeness of formal methods by refining a disturbance model using online collected data and it evaluates the safety of a learning-based controller, using computationally efficient zonotopic reachability analysis for the safety analysis to facilitate a real-time implementation. We validate the framework in a real-world drone flight through a canyon, where the drone is subjected to unknown external disturbances and the framework is tasked with learning those disturbances online and adjusting the safety guarantees accordingly. The results show that the framework enables a less restrictive online training of learning-based controllers without compromising the safety of the system.

\end{abstract}

\section{Introduction \& Related Work}
\label{sec:introduction}

Reinforcement Learning (RL) demonstrates significant potential in robotic and autonomous systems, particularly for handling complex tasks in uncertain environments. While RL's data efficiency and policy optimality are well studied~\cite{Singh2022Reinforcement}, its safety and reliability are pressing research for learning on real hardware. Although safety in RL has been discussed for decades~\cite{HegerConsideration1994}, it has only recently begun to take on a more concrete form~\cite{GarciaComprehensive2015} and its provision remains an open problem.
One area actively explored to tackle this challenge is Provably Safe RL (PSRL). As discussed in~\cite{FisacGeneral2019, SelimSafe2022, KochdumperProvably2023, Krasowski2023Provably}, they employ shielding mechanisms to evaluate the safety of an agent's action by predicting potential future outcomes. The action is either deemed safe or unsafe in which case it is replaced with a provably safe alternative action accompanied by some penalty for its unsafe choice. While PSRL can offer strong safety guarantees, its solutions often evaluate action safety based on pre-computed safe sets, or they may use conservative worst-case assumptions, such as an a priori assumed maximum external disturbance vector~\cite{Gillula2011Guaranteed}. Such restrictive assumptions, combined with penalizing unsafe choices~\cite{kochdumper2023provably}, can discourage the agent from exploring certain state/action pairs. Although, this promotes safety, overly conservative assumptions can unnecessarily limit exploration. Addressing this, is imperative as the hindrance to the RL exploration can lead to sub-optimal performance or even failures on task completion. 
A natural response to this restrictiveness is to develop adaptive methods that refine the assumptions online~\cite{shi2019neural, berkenkamp2017safe}. Specifically, in this work, we focus on uncertainties modeled as disturbances to the system's state. 
An existing approach addressing this was introduced in~\cite{FisacGeneral2019} and was later extended in~\cite{Akametalu2014Reachability}, where the authors employed Hamilton-Jacobi-Isaacs (HJI) reachability analysis (RA) to infer the possible future outcomes for the system. In addition they used a Bayesian approach to model the disturbance function as being drawn from a Gaussian process dependent on the system's current state. 
%
Although the works in~\cite{Akametalu2014Reachability, FisacGeneral2019} allow for online disturbance adaptations they suffer from the curse of dimensionality as they use HJI RA. RA, is used to reevaluate safety under the updated disturbance model. To do so it generally involves approximating all possible system trajectories and then determine if a trajectory intersects a denoted unsafe region in the sapce. This makes HJI generally limited to low dimensinoal systems despite of being a powerful and general approach ~\cite{Bansal2017Hamilton}. 

The real-wrold applicability of RL methods is of paramount importance. As such, employing HJI, or not utilizing online collected data about external disturbances to refine the system's view of the world can be a significant bottleneck on some of the vital perks of RL; its ability to handle highly complex systems and exploring previously unknown strategies.
In turn, we present a framework that integrates online disturbance learning with efficient RA for less restrictive exploration with the potential of uncompromised safety, even when disturbance patterns change during the mission. We employ a stochastic gradient descent (SGD) regressor to predict future disturbance bounds based on a recent history of state-action-disturbance triplets, capable of identifying disturbance patterns along the system's trajectories. The regressors' continuously refined estimates are then used to adapt the safety constraints accordingly. To address the HJI's scalability issues like in~\cite{FisacGeneral2019}, we use a zonotopic RA method~\cite{Girard2005Reachability, Yang2022Efficient, Bird2023Hybrid}. Finally, we devise a shielding mechanism to replace unsafe policy actions with empirically well-performing or even provably safe alternatives.
This approach offers a less restrictive solution for online training of learning-based controllers without compromising system safety. Our contributions are: (1) a refined disturbance model for online, on-policy training in the context of learning-based control with safety guarantees, (2) the online computation of safety sets using zonotopic RA, and (3) a successful evaluation of the framework on consumer-level hardware, reducing the restrictiveness of the agent's training process.

\section{Preliminaries}
\label{sec:preliminaries}


\noindent Consider the discrete-time nonlinear system
\begin{align}
    \vect{x}_{t+1} = f(\vect{x}_{t}, \vect{u}_{t}, \vect{w}_{t}),
    \label{eq:prelim:dynamics_model}
\end{align}
where $\vect{x}_{t} \in \mathbb{R}^{n_{x}}$ is the state vector, $\vect{u}_{t} \in \mathbb{R}^{n_{u}}$ is the control input, $\vect{w}_{t} \in \mathbb{R}^{n_{x}}$ is the disturbance, and $f: \mathbb{R}^{n_{x}} \times \mathbb{R}^{n_{u}} \times \mathbb{R}^{n_{x}} \rightarrow \mathbb{R}^{n_{x}}$ the dynamics of the system. For each time instant $t$, $\vect{u}_{t} \in \mathbb{U} \subset \mathbb{R}^{n_{u}}$, and $\vect{w}_{t} \in \mathbb{W} \subset \mathbb{R}^{n_{x}}$, where $\mathbb{U}$ and $\mathbb{W}$ are the compact control input and disturbance constraints respectively. The system is assumed to be well-defined so that $\vect{x}_{t+1}$ is uniquely determined by $\vect{x}_t$, $\vect{u}_t$ and $\vect{w}_{t}$. Let $\eta = \vect{u}_0, \vect{u}_1, \ldots$ be a control policy and $\mathcal{H}$ be the set of all control policies, as well as $d = \vect{w}_{0}, \vect{w}_{1}, \ldots$ be a sequence of disturbances and $\mathcal{D}$ the set of all disturbance trajectories. For initial state $\vect{x}_0$, we denote a trajectory of~\eqref{eq:prelim:dynamics_model} as $\zeta(\cdot; \vect{x}_0, \eta, d)$ where $\zeta(t; \vect{x}_0, \eta, d)$ is the state of system at time $t$ starting from $\vect{x}_0$ and applying policy $\eta$ under disturbance $d$. 

\subsection{Reachability Analysis}
\label{subsec:reachability_analysis}

RA is a key tool for providing formal guarantees on system behavior. In this work, it's used to assess control input safety by evaluating the safety of all reachable states under each input for all disturbance levels. Thus, the most relevant RA definitions here are finite-horizon backward and forward reachable sets (BRS, FRS).

\vspace{0.35cm}
\begin{definition}
       Given~\eqref{eq:prelim:dynamics_model}, the finite time $N$-step BRS from the target set $\mathcal{T} \subset \mathbb{R}^{n_{x}}$ within the space $\mathbb{S} \subset \mathbb{R}^{n_{x}}$ is given as
       \begin{equation}
        \begin{split}
            \mathcal{R}_{b}(\mathcal{T}, N) = 
            \{
            & \vect{x} \in \mathbb{S}| \exists \eta \in \mathcal{H}, \forall d \in \mathcal{D}, \\   & \exists t \in [0, N], \zeta(t; \vect{x}, \eta, d) \in \mathcal{T}
            \}.
        \end{split}
        \label{eq:brs}
       \end{equation}
\end{definition}
\vspace{0.35cm}

Intuitively the N-time BRS answers the question of where can the system be and eventually attain a state in region $\mathcal{T}$ in $N$ units of time. In practice, to compute the BRS one computes the union of the predecessor sets for all time steps $t \in [0, N]$. Fundamentally, the predecessor set of $\mathcal{T}$ is the set of all states that can lead to $\mathcal{T}$ in a single time step and can be defined as
\begin{equation}
    \begin{split}
        Pre(\mathcal{T}) \!=\! 
        \{& \vect{x} \in \mathbb{S} | \exists \vect{u} \! \in \! \mathbb{U}, \forall \vect{w} \in \mathbb{W}, f(\vect{x}, \vect{u}, \vect{w}) \in \mathcal{T}
        \}.
    \end{split}
    \label{eq:predecessor_set}
\end{equation}

Then the N-step BRS is equivalent to computing the union of $N$ predecessor sets and can for example be computed as $\mathcal{R}_{b}(\mathcal{T}, 2) = Pre(\mathcal{T}) \cup Pre(Pre(\mathcal{T})).$
%

The FRS tells us which states can the system attain in the future starting from a target region $\mathcal{I} \in \mathbb{R}^{n_{x}}$.
\begin{definition}
       Given~\eqref{eq:prelim:dynamics_model}, the finite time $N$-step FRS from the initial set $\mathcal{I} \subset \mathbb{R}^{n_{x}}$ within the space $\mathbb{S} \subset \mathbb{R}^{n_{x}}$ is given as
       \begin{equation}
        \begin{split}
            \mathcal{R}_{f}(\mathcal{I}, N) = 
            \{
            & \zeta(t; \vect{x}, \eta, d) \in \mathbb{S} | \vect{x} \in \mathcal{I}, \\
            & \exists \eta \in \mathcal{H}, \exists d \in \mathcal{D}, t \in [0, N]
            \}.
        \end{split}
        \label{eq:frs}
       \end{equation}
\end{definition}
\vspace{0.15cm}
Similarly we compute the union of the successor sets for all time steps $t \in [0, N]$. which are in turn defined as

\begin{align}
    Suc(\mathcal{I}) \!=\! 
    \{
    f(\vect{x}, \vect{u}, \vect{w}) \in \mathbb{S} | \vect{x} \in \mathcal{I}, \exists \vect{u} \in \mathbb{U}, \exists \vect{w} \in \mathbb{W}
    \}.
    \label{eq:successor_set}
\end{align}

On a closer look, the RA definitions~\eqref{eq:brs} to \eqref{eq:successor_set} hint the effect the disturbance model on the resulting sets and on the safety guarantees. For instance larger bounds on the set $\mathbb{W}$ can result in a bigger FRS and a tighter BRS.

\subsection{Zonotopes \& Zonotopic Reachability Analysis}
\label{subsec:zonotope}
The zonotope is a centrally symmetric convex polytope defined as the affine image of a unit hypercube and has been shown to handle approximations of reachable sets in a computationally efficient manner~\cite{Yang2021Scalable}.
\begin{definition}
    A set $\mathcal{Z} \subset \mathbb{R}^{n}$ is a zonotope if $\exists G \in \mathbb{R}^{n \times n_{g}}$ and $\exists \vect{c} \in \mathbb{R}^{n}$, where $n, n_{g}$ are the space dimensionality and number of generators the zonotope has respectively, such that $\mathcal{Z} =\{ \vect{c} + G \vect{\xi} ~|~ ||\vect{\xi}||_{\infty} \leq 1 \}.$
\end{definition}

For RA, zonotopes, require that the system dynamics are linearized in the form of $\vect{x}_{t+1} = A\vect{x}_{t} + B\vect{u}_{t}+\vect{w}_{t}$, where $A \in \mathbb{R}^{n_{x} \times n_{x}}, B \in \mathbb{R}^{n_{x} \times n_{u}}$ are the discretized state transition and control input matrices. Then, we compute the predecessor~\eqref{eq:predecessor_set} and successor~\eqref{eq:successor_set} sets using the formulas derived from Minkowski arithmetic~\cite{Kurzhanskiy2011Reach} as follows

\vspace{-0.35cm}
\begin{align}
    & Pre(\mathbb{S}, \mathcal{T}) = A^{-1}(\mathcal{T} \ominus \mathbb{W} \oplus -B\mathbb{U} )\label{eq:predecessor_linear_systems}, \\
    & Suc(\mathbb{S}, \mathcal{I}) = A\mathcal{I} \oplus B\mathbb{U} \oplus \mathbb{W},\label{eq:successor_linear_systems}
\end{align}

where $\oplus$ and $\ominus$, are the Minkowski sum and Minkowski difference respectively which in this context are defined as $\mathbb{A} \oplus \mathbb{B} = \{\vect{a}+\vect{b} ~|~ \vect{a} \in \mathbb{A}, \vect{b} \in \mathbb{B} \}$ and $\mathbb{A} \ominus \mathbb{B} = \{\vect{c} \in \mathbb{R}^{n} ~|~ \exists \vect{b} \in \mathbb{B}: \vect{c} +\vect{b} \in \mathbb{A} \}$. For both the predecessor and successor sets~\eqref{eq:predecessor_linear_systems}, \eqref{eq:successor_linear_systems} the sets $\mathcal{T}, \mathcal{I}$ are a zonotope infinitesimally small around the current state $\vect{x}_{k}$.

\subsection{Action safety evaluation and replacement}

The proposed framework requires that the safety of a state-action pair can be verified. To do so, one can define a safety function~\cite{Krasowski2023Provably} $\varphi: \mathbb{R}^{n_{x}} \times \mathbb{R}^{n_{u}} \rightarrow \{0, 1\}$ as

\vspace{-0.2cm}
\begin{align}
    \varphi(\vect{x}, \vect{u}) = 
    \begin{cases}
        1, & \text{if} ~(\vect{x}, \vect{u}) ~\text{is safe,} \\
        0, & \text{otherwise.}
    \end{cases}
    \label{eq:safety_function}
\end{align}

\begin{figure}[H]
    \centering
    \includegraphics[width=0.37\textwidth, trim=0.7cm 0.33cm 0.8cm 0.45cm, clip]{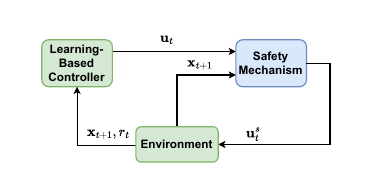}
    \caption{Replacement of unsafe action by a provably safe or empirically well-behaved one.}
    \label{fig:action_replacement_mechanism}
\end{figure}

The value of the safety function $\varphi(\cdot)$ depends on the precise notion of safety which we shall define later in the next section as a function of the reachable sets introduced earlier. 
The utility of $\varphi(\cdot)$ is in that it allows us to now evaluate whether an action chosen by the trained controller is acceptable or if it might emerge prohibitive behavior. In the latter case the action needs be replaced by an alternative one that is either provably safe, or simply a well behaved one and generally denoted by $\vect{u}_{t}^{s}$. This is a common setup in the realm of safe RL and is visually depicted in Figure~\ref{fig:action_replacement_mechanism}.

\section{Methodology}
\label{sec:methodology}

To enable less restrictive yet safe training of learning-based controllers for complex robotic systems in uncertain conditions, we propose an adaptive method that integrates SGD regression models with RA. This method includes two key components alongside the RL agent and the environment: the \textit{Disturbance Learning Unit} and the \textit{Safety Mechanism}. 

The interplay amongst these units is illustrated in Figure~\ref{fig:framework_block_diagram}. Specifically, for the replacement action we opt for the scheme

\begin{align}
    \vect{u}_{t}^{s} = 
    \begin{cases}
        \vect{u}_{t}, & \text{if} ~\varphi(\vect{x}, \vect{u}_{t}) = 1, \\
        \vect{u}_{t}^{r}, & \text{otherwise},
    \end{cases}
\end{align}
\vspace{0.2cm}

where $\vect{u}_{t}^{r}$ is the backup action, e.g. chosen by a classic controller. At the same time, the agent interacts with an environment of both controllable (the agent can directly influence) elements governed by~\eqref{eq:prelim:dynamics_model}, and uncontrollable ones (e.g., obstacles). In parallel, the disturbance learning unit, makes a prediction on the disturbance levels the plant is expected to experience. Combining this prediction with the policy action, and some spatial awareness (e.g., obstacles' position), the safety mechanism evaluates the safety of the chosen action. If deemed unsafe, the action replacement method is activated and the agent skips its training step for that time step.

\vspace{0.2cm}
\subsection{Learning-Based Control}
\label{subsec:methodology:learning_based_control}

The policy used in this work is implemented by a Gaussian multilayer perceptron (MLP) which takes as input the state error vector $(\vect{x}_{t} - \vect{g}_{t}) \in \mathbb{R}^{n_{x}} $, where $\vect{g}_{t}$ is the goal state, and outputs the probabiliy distributions $\mathcal{N}_{i}(\mu_{i}, \sigma), ~i = [1, n_{u}]$ for the control vector.

The network's parameters are optimized using a policy gradient approach, suited for continuous state and action spaces. The objective is to minimize a loss function $L(\theta)$ following the update rule $\vect{\theta}_{t} = \vect{\theta}_{t-1} + \alpha\nabla_{\vect{\theta}} L(\vect{\theta}),$
where $\alpha$ is the learning rate and the loss function is defined as $L(\vect{\theta}) = || \vect{u}^{r}_{t}(\vect{x}_{t}, \vect{g}_{t}) - \vect{u}_{t}||_{2}$.

\vspace{0.2cm}
\subsection{Disturbance Learning Unit}
\label{subsec:disturbance_learning_unit}

The a priori assumed disturbance model provides a worst-case estimate of the disturbance's impact on the system. This overlooks that higher-order dynamics also abide by physics' laws and do not change arbitrarily. So, we aim to learn how the disturbance evolves over time. Thus, the purpose of this unit is to improve predictions of future disturbances by training online based on the actual disturbances the system encounters. The disturbance assumed in this work refers to any effects the $\vect{w}_{t}$ component of~\eqref{eq:prelim:dynamics_model} has and if assumed to be additive it can be computed as $\vect{w}_{t} = \vect{x}_{t+1} - f(\vect{x}_{t}, \vect{u}_{t}, \vect{0})$, where $f(\vect{x}_{t}, \vect{u}_{t}, \vect{0})$ are the low order dynamics assumed to be known to us.

During the system's operation a log of experiences is maintained. An experience is a tuple consisting of the system's state, the control input it chose to take, as well as the disturbance it experienced at that time step under that state-input pair. Given this memory, a history of state, action, disturbance triplets $(\vect{x}_{t}, \vect{u}_{t}, \vect{w}_{t})$ is extracted to train an SGD regressor model to learn a probability distribution of the disturbance sequence $d$ the system expects to experience in the near future; that is the next horizon $(N)$ time steps. 

The score function of the regressor is given by $f(S_{I}) = w^{T}S_{I} + b$ with model parameters $w \in \mathbb{R}^{n_{o} \times n_{i} }$ and intercept $b \in \mathbb{R}^{n_{0}}$, where $n_{i}$, $n_{0}$ are the dimensions of the input and output samples given by $n_{i} = (n_{x} + n_{u})N + n_{x} + 2$, $n_{o} = 2$. The model parameters, are derived from minimizing the regularized training error

\vspace{-0.15cm}
\begin{equation}
    E(w, b) = 
    \frac{1}{n} \sum_{i=1}^{n} L_{SGD}(S_{O}^{i}, f(S_{I}^{i})) + aR(w),
\end{equation}
\vspace{0.1cm}

where $n$ is the number of samples and $L$ is the loss function

\vspace{-0.2cm}
\begin{equation}
    L_{SGD}(S_{O}^{i}, f(S_{I}^{i})) = \frac{1}{2} ||S_{O}^{i} - f(S_{I}^{i})||.
\end{equation}
\vspace{0.1cm}

Regarding the input data samples for the regressor they have the following structure

\begin{figure}[t]
    \centering
    \vspace{0.3cm}
    \includegraphics[width=0.5\textwidth, trim=5.4cm 14.25cm 7.1cm 7.68cm, clip]{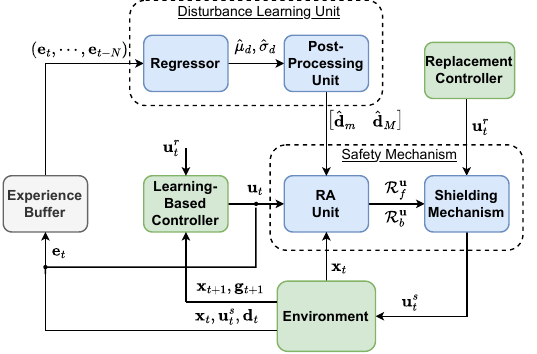}    
    \caption{Proposed Framework - Online disturbance learning and reevaluation of safety guarantees to enable the safe training of a learning-based controller.}    
    \label{fig:framework_block_diagram}
    \vspace{-0.5cm}
\end{figure}

\begin{equation}
    \label{eq:regressor_sample_input}
    \begin{split}
        & S_{I} = 
        \begin{bmatrix}
            \vect{x}_{h} & \vect{u}_{h} & \mu_{d} & \sigma_{d}
        \end{bmatrix}, \\[0.3cm]
        & \vect{x}_{h} = 
        \begin{bmatrix}
            \vect{x}_{t-N}^{T} & \vect{x}_{t-(N-1)}^{T} & \cdots & \vect{x}_{t}^{T}
        \end{bmatrix}, \\[0.3cm]
        & \vect{u}_{t} =
        \begin{bmatrix}
            \vect{u}_{t-N}^{T} & \vect{u}_{t-(N-1)}^{T} & \cdots & \vect{u}_{t-1}^{T}
        \end{bmatrix}, \\[0.3cm]
        & \mu_{d} = \sum_{i=0}^{N-1}\frac{\vect{w}_{t-i}}{N}, \hspace{0.3cm}
        \sigma_{d} = \sqrt{\frac{\sum_{i=0}^{N-1}(\vect{w}_{t-i}-\mu_{d})^{2}}{N}}.
    \end{split}
\end{equation}
\vspace{0.2cm}

Similarly, the output of the regressor consists of a predicted mean value and a standard deviation. For instance, during the training a sample output fed to the regressor has the following structure

\begin{equation}
    \label{eq:regressor_sample_output}
    \begin{split}
        & S_{O} = 
        \begin{bmatrix}
            \hat{\mu}_{d} & \hat{\sigma}_{d}
        \end{bmatrix}, \\
        & \hat{\mu}_{d} = \sum_{i=0}^{N-1}\frac{\vect{w}_{t+i}}{N}, \hspace{0.3cm}
        \hat{\sigma}_{d} = \sqrt{\frac{\sum_{i=0}^{N-1}(\vect{d}_{t+i}-\mu_{w})^{2}}{N}}.
    \end{split}
\end{equation}
\vspace{0.1cm}

Given the regressor's output $S_{O} = \begin{bmatrix} \hat{\mu}_{d} & \hat{\sigma}_{d} \end{bmatrix}$, the predicted upper and lower bounds of the constrained disturbance $\mathbb{W}$ from which all possible disturbance sequences $d$ can be constructed from are given as $[\hat{\vect{d}}_{m}, \hat{\vect{d}}_{M}] = [\vect{\hat{\mu}}_{d} - 3\vect{\hat{\sigma}}_{d}, \vect{\hat{\mu}}_{d} + 3\vect{\hat{\sigma}}_{d}]$.
This implies that we assume the environmental disturbances can be captured to a good enough extent using a Gaussian distribution. However, the proposed framework is not strictly limited to this choice, as long as the true disturbance under consideration is well-behaved. For instance, a human abruptly pushing the drone mid-flight is not natively supported by the proposed framework. Whereas, for well-behaved disturbances, such as wind gusts the regressor can be augmented to include higher order moments. In turn, the post-processing unit can be adjusted to reflect the new disturbance bounds instead of the $\pm 3$ standard deviations.

\vspace{0.1cm}
\subsection{Shielding Mechanism}
\label{subsec:contorl_input_shielding}

Moving on to the safety mechanism, we first need to define the notion of safety for an action which will in turn allow us to evaluate the safety function~\eqref{eq:safety_function}. Given the current state $\vect{x}_{t}$, an action is deemed safe if it is an element of the locally safe action space defined as

\vspace{-0.15cm}
\begin{equation}
    \begin{split}
        \mathcal{U}_{\vect{x, t}}^{s} = 
        \{ \vect{u} \in \mathbb{U} |
        &\mathcal{R}_{f}^{u}(\vect{x}, N) \cap \mathcal{O}bs = \emptyset, \\
        &\mathcal{R}_{f}^{u}(\vect{x}, N) \subseteq \mathcal{R}_{b}(\vect{x}, N)\},
    \end{split}
    \label{eq:safe_action_space}
\end{equation}
\vspace{0.1cm}

where $\mathcal{R}_{f}^{u}(\mathbb{S}, \vect{x}, N)$ is the forward reachable set but with the action space confined to be the action under evaluation instead. In addition $\mathcal{O}bs$ denotes any nonsafe regions, such as physical obstacles in the environment. Therefore, that safety function~\eqref{eq:safety_function} will have a value of $1$ whenever $\vect{u}_{t} \in \mathcal{U}_{\vect{x, t}}^{s}$, otherwise it will be $0$.

This notion of safety consists of two parts. The first, aims to ensure the system is not in immediate danger of entering an unsafe region. The latter, evaluates whether the system can recover to its current state if needed, thus, we check if the BRS from the current state fully contains the FRS. Although, at first the second safety check might seem untrustworthy, since it relies on the current state to be safe, we can see that it is indeed reliable if we consider that this process runs from the launch time and as such the system will not allow the agent in the first place to reach an unsafe state. Finally, the action replacement unit, will allow the chosen action $\vect{u}_{t}$ to pass through if $\varphi(\vect{x}_{t}, \vect{u}_{t}) = 1$, otherwise, it will replace it with an action $\vect{u}_{t}^{r}$ from the replacement controller as shown in Figure~\ref{fig:framework_block_diagram}.

\vspace{0.5cm}
\section{Drone Flight Through Canyon Scenario}
\label{sec:drone_flight_through_canyon_scenario}

In this section, we assess our method experimentally in a real-world scenario. A compelling use case is navigating a drone through narrow and confined spaces such as a narrowing canyon. Flying a quadrotor in such an environment introduces a number of challenges, such as the reflection of the thrusted air from the walls and the ground as well as the overall reduced space for maneuvering. 

\begin{figure}[b]
    \centering
    \vspace{0.2cm}
    \includegraphics[width=\linewidth]{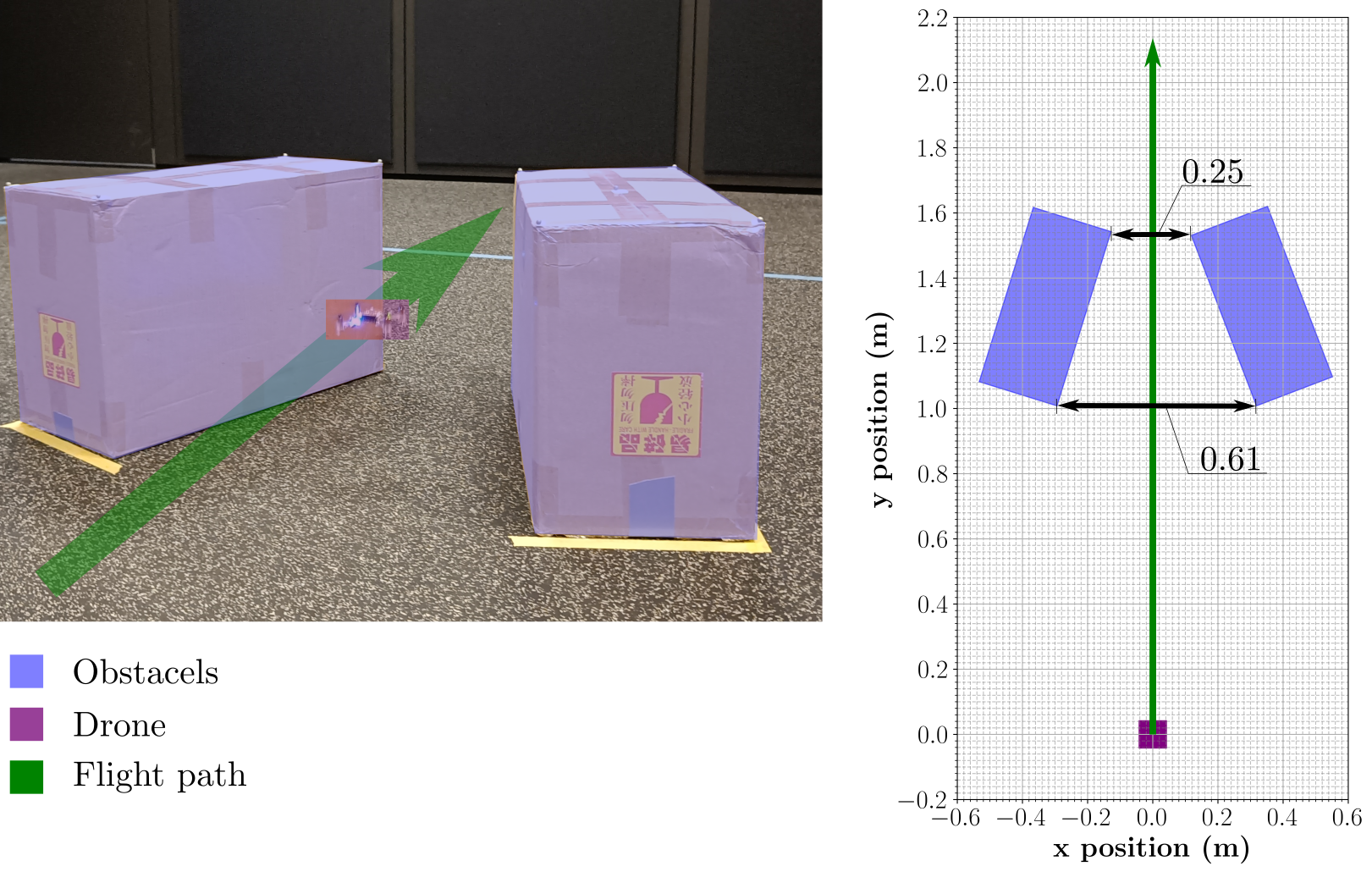}
    \caption{Canyon Environment in the lab - showing the lab setup on the left and a rendered top view on the right.}    
    \label{fig:canyon_environment}
\end{figure}

Figure~\ref{fig:canyon_environment} shows the narrowing canyon setup in our lab (left) and a plot as a top view (right). 
As the drone follows the flight path the canyon gets progressively narrower making the flight increasingly more challenging as obstacle closeness and increased air disturbances (from the walls) occur. 
We are interested in using the scenario to assess and answer questions as follows:
\begin{itemize}
    \item Can the proposed learning unit predict the pattern of the disturbance variation?
    \item Can the disturbance prediction be used by the zonotope-based analysis to yield less restrictive safe sets?
    \item Can the derived safe sets decrease the probability of unnecessarily rendering the system unsafe and thus allow the training agent for more data exploration?
\end{itemize}

The last item can be assessed through the number of times the agent's action has been deemed safe in the same training episodes, which in turn links to the training exposure the learning-based controller receives. 
To this end, we compare the performance of the proposed framework with a restrictive adaptation of it where the disturbance model is not refined online and instead the a priori assumed worst-case disturbance bounds are used and the safety of the drone can be completely determined a priori based on those assumptions~\cite{Gillula2011Guaranteed}.

\begin{figure*}[t]
    \centering
    \includegraphics[width=\textwidth, height=0.55\textwidth]{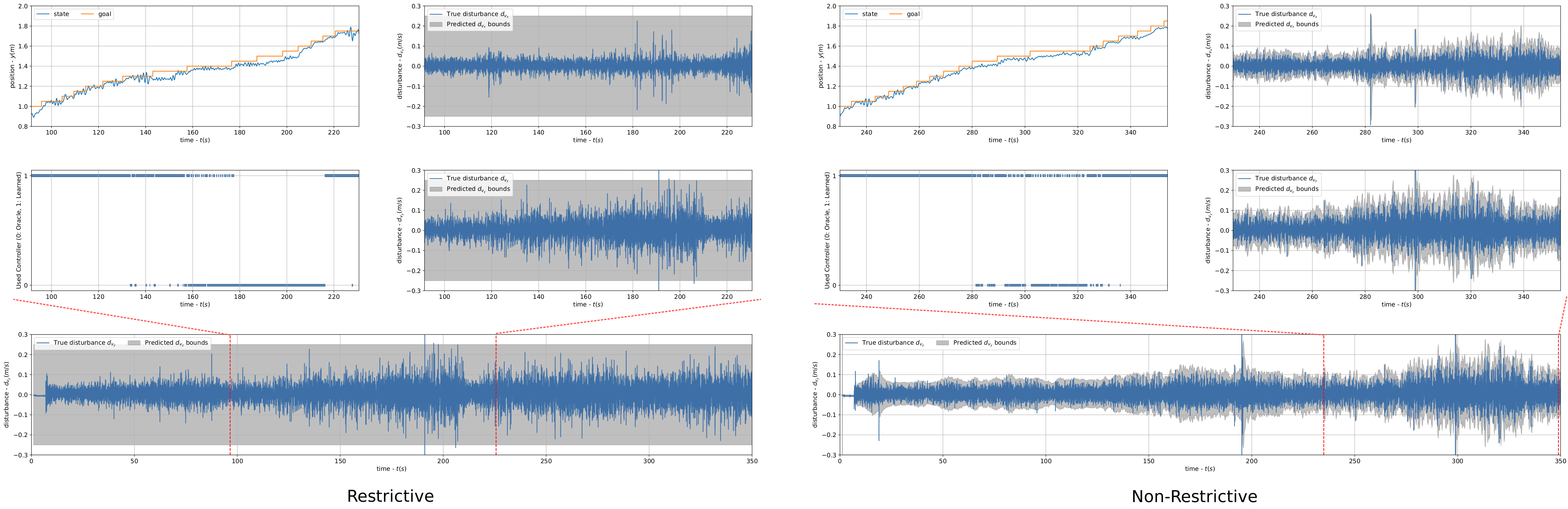}
    \vspace{0.2cm}
    \caption{Quantitative view of the impact from non-restrictive online disturbance learning Vs restrictive a priori assumed disturbances on the training of the learning-based controller.}    
    \label{fig:conservative_vs_nonconservative}
\end{figure*}
\vspace{0.2cm}
\subsection{Drone Model}
\label{subsec:drone_flight:drone_model}

The drone used in this mission is a Crazyflie 2.1~\cite{Giernacki2017Crazyflie}, a small quadrotor equipped with an onboard controller. The platform offers the option of a cascaded controller, where the inner loop which regulates the orientation of the drone is already implemented onboard. The developed framework, is instead focused on the outer loop dealing with the position controller. As such, the dynamics of the drone can be given in a state-space representation as

\vspace{0.01cm}
\begin{equation}
\begin{split}
    \dot{\begin{bmatrix}
        x \\
        y \\
        z \\
        v_{x} \\
        v_{y} \\
        v_{z}
    \end{bmatrix}}
     = 
    \underbrace{\begin{bmatrix}
        v_{x} \\
        v_{y} \\
        v_{z} \\
        \frac{f_{T}}{m} cos(\phi) sin(\theta) cos(\psi) + sin(\phi) sin(\psi) \\
        \frac{f_{T}}{m} cos(\phi) sin(\theta) sin(\psi) + sin(\phi) cos(\psi) \\
        \frac{f_{T}}{m} cos(\phi) cos(\theta) - g
    \end{bmatrix}}_{f(\vect{x}, \vect{u})}
    +
    \vect{w}.
\end{split}
\label{eq:drone_nonlinear_dynamics}
\end{equation}
\vspace{0.2cm}

The state vector of the plant~\eqref{eq:drone_nonlinear_dynamics} consists of the position of the drone's center of mass $(x, y, z)$ in global coordinates and its translational velocity along these axis $(v_{x}, v_{y}, v_{z})$. The control input vector is given as $\vect{u} = \begin{bmatrix} f_{T}, \phi, \theta \end{bmatrix}$, where $f_{T}$ is the collective thrust force from all four motors, $\phi, \theta$ are the roll and pitch Euler angles respectively. The mass of the drone and the acceleration of gravity are denoted by $m, g$ respectively.

The zonotopic reachability analysis requires linearized dynamics. Linearization can introduce modelling errors which in turn might render any guarantees invalid. For instance, in this case, the analysis should in principle only hold for limited angular motions of the drone. However, we can bypass this problem by adjusting the bounds of the constraints of the control input set and state space to account for the effect of the pruned non-linearities.
Note that other methods such as conservatively linearizing the dynamics of the system using Lagrangian remainders~\cite{Yang2022Efficient} are also possible. Using the small angle approximation we get $\dot{\vect{x}} = A\vect{x} + B\vect{u} + \vect{w}$, 
where the continuous-time $A$ and $B$ matrices are given as

\begin{equation}
    \begin{split}
    A = 
    \begin{bmatrix}
        0 & 0 & 0 & 1 & 0 & 0 \\
        0 & 0 & 0 & 0 & 1 & 0 \\
        0 & 0 & 0 & 0 & 0 & 1 \\
        0 & 0 & 0 & 0 & 0 & 0 \\
        0 & 0 & 0 & 0 & 0 & 0 \\
        0 & 0 & 0 & 0 & 0 & 0
    \end{bmatrix}    
    B = 
    \begin{bmatrix}
        0 & 0 & 0 \\
        0 & 0 & 0 \\
        0 & 0 & 0 \\
        0 & 0 & g \\
        0 & -g & 0\\
        \frac{1}{m} & 0 & 0
    \end{bmatrix}.
    \end{split}
\end{equation}
\vspace{0.2cm}


\subsection{Experimental Setup}
\label{subsec:drone_flight:infrastructure}

The off-board controller is implemented on an Ubuntu 20.04 LTS machine with an AMD Ryzen 7 4700u processor. The communication amongst the different components of the framework in Figure~\ref{fig:framework_block_diagram} is facilitated through the Robot Operating System (ROS) Noetic.  It is assumed that real-time ground truth information about the drone's state is provided by a Motion Capture (MoCap) system. The Crazyflie python API\footnote{https://github.com/bitcraze/crazyflie-lib-python} is used to link the ROS implementation of the off-board controller with the onboard low level controller on the Crazyflie. The drone is placed at $(0,0)$ as indicated in Figure~\ref{fig:canyon_environment} (right). After take-off the drone follows the flightpath by incrementing its goal position in the $y-$axis in steps of $10cm$ at the beginning of the flight up until the canyon entrance $(y<1m)$ and steps of $5cm$ from the canyon entrance and onwards. The goal updates are given every $10$ seconds with the additional restriction that the drone has to first stabilize well enough around the desired state before the next goal position is sent. Specifically, we experimentally determined that it suffices for the position and velocity errors to be less than $0.075m$ and $0.1m/s$ respectively for the last 25 time steps (half a second). As for the replacement controller, a linear quadratic regulator (LQR) designed according to the undisturbed version of the linearized dynamics.

\vspace{0.2cm}
\section{Results}
\label{sec:results}

\begin{figure}[t]
    \centering
        \includegraphics[width=\linewidth]
        {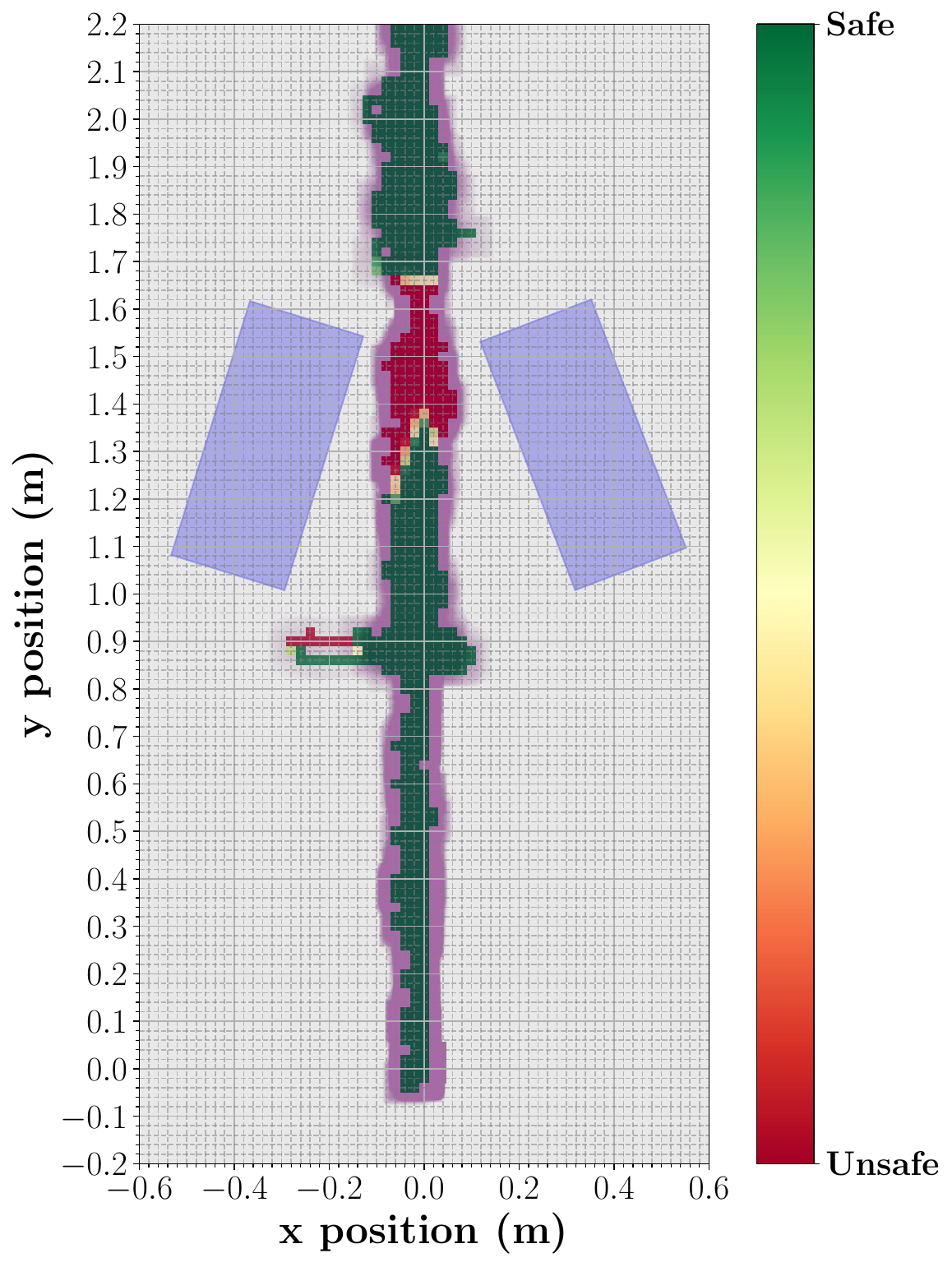}
        \vspace{0.2cm}
        \caption{Aggregated results of five flights for Restrictive setting.}
        \vspace{-\baselineskip}
        \label{fig:state_trajectory_conservative}
        \vspace{0.15cm}
\end{figure}

\begin{figure}[t]
        \includegraphics[width=\linewidth]{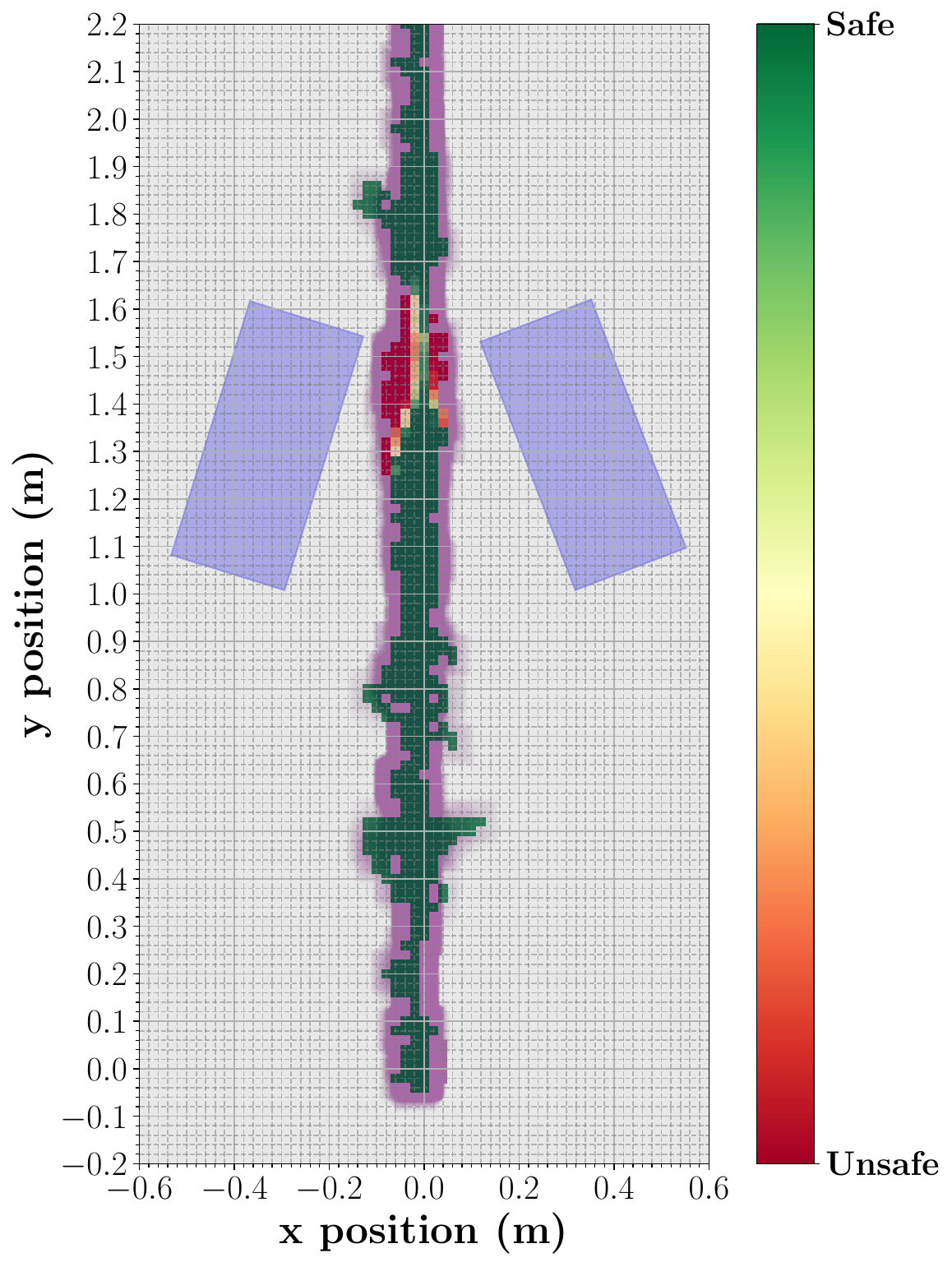}
        \vspace{0.2cm}        
        \caption{Aggregated results of five flights for Non-Restrictive setting.}
        \vspace{-\baselineskip}
        \label{fig:state_trajectory_nonconservative}
        \vspace{0.1cm}
\end{figure}

Starting with a quantitative evaluation of the framework, the results are portrayed in Figure~\ref{fig:conservative_vs_nonconservative}, where the training process of the agent under a restrictive and non-restrictive flight is compared. On the left, we see the former, where the DLU essentially operates like a min-max operator, driving its output to some a priori assumed worst-case upper and lower disturbance levels as indicated by the shaded gray area in all three disturbance subplots.
Starting with the plot at the bottom of the restrictive side (left), the evolution of the disturbance throughout the entire trajectory is displayed alongside the conservative bounds. 
Immediately noticeable is the fact that the disturbance levels are higher when the drone is well deep inside the canyon. This fact as well as the proximity to the walls around that period suggest that the system will be more active in preventing any unwanted actions chosen by the trained controller. To better study the frameworks behavior, we showcase the period the drone is within the canyon in the four plots above.
The top left figure shows the position in y-direction indicating how the drone is following the straight flight path. We can observe how the incremental goal-points of 5cm are sent once the conditions discussed before are fulfilled. The two figures on the right of it show the true disturbance in $d_{vx}$ and $d_{vy}$. For both dimensions we can see that the disturbance generally increases towards the end, and therefore narrower, part of the canyon.
The zoomed in left bottom figure shows at what time the drone has deemed a learned action as unsafe and had to revert back onto its fallback LQR controller. We can see that the method judged a large consecutive number of actions as unsafe once the more narrow part of the canyon was reached ($1.4 \leq y \leq 1.6 m$) and even slightly after it exited the canyon while $y \leq 1.70m$. We can contrast with our less restrictive method shown on the right of Figure~\ref{fig:conservative_vs_nonconservative}. The general trend of the disturbance (shown in the bottom right) is similar as the drone follows the same flight path, the core difference lies in the number of times our method had to rely on the LQR. We can see that while some actions had to be replaced by the oracle, in contrast to the restrictive version the replacement was much more sporadic and not consecutive. This suggests that the learning based controller can acquire many more new data that can be used to further train the system.
Notably, the disturbance plots for $d_{vx}$ and $d_{vy}$ highlight how the disturbance learning unit sufficiently captures the true inflicted disturbance as well as how a tighter bound on the predicted disturbances can result in increased performance.

To mitigate the naturally stochastic nature of the experiments, we performed five flights with each variation. We report the aggregated results in Figures~\ref{fig:state_trajectory_conservative},~\ref{fig:state_trajectory_nonconservative}, where we discretize the space into a two-centimeter grid and color each cell according to the ratio of safe states over the total states we record (safe plus unsafe states), If a grid cell is not visited by the drone's center we color it gray, The drone itself is plotted as a purple square around each drone center as they are considered in the zonotopic RA.
When analyzing Figure~\ref{fig:state_trajectory_conservative} we can observe that most actions suggested at locations near the narrow end of the canyon are deemed unsafe while our approach shown in Figure~\ref{fig:state_trajectory_nonconservative}
triggers the safety replacement in much fewer cases.
This highlights the potential of our approach as it can evidently provide a less restrictive solution for safe online training of learning-based time-critical systems.
Indeed when aggregating the safe-to-unsafe ratio of all non-empty grid cells in the canyon region ($y=[1.2,1.7], x=[-0.4,0.4] $) and then dividing it by the number of non-empty cells
we obtain only $24.99\%$ safe actions for the restrictive disturbance unit while our non-restrictive approach manages to have $63.1\%$ on average in the mentioned region.

\vspace{0.2cm}
\section{Conclusion}
\label{sec:conclusion}

This paper presents a new framework for less-restrictive training of learning-based controllers by adapting online its assumed disturbance online based on real world data and adjusting its safety constraints accordingly. The solution utilizes the learning capabilities of SGD regressors to refine its disturbance model and a zonotopic RA for evaluating the safety constraints. The applicability of the proposed framework is demonstrated in a real-world drone flight through a canyon scenario under unknown external disturbances. The results show that the proposed framework can provide a less restrictive training ground for learning-based controllers with the potential to not compromise the safety of the system. Future work will focus on exploring the applicability of the framework to more intricate RL controllers under a variety of disturbance sources as well as evaluate its potential for increased performance as a result of the less restrictive training environment.

\section*{Acknowledgments}
This work was supported by the Swedish Research Council, the Knut and Alice Wallenberg Foundation, the European Research Council (ERC Grant No. 884807). The authors gratefully acknowledge this support.

\vspace{0.5cm}

\bibliography{paper}
\bibliographystyle{IEEEtran}

\end{document}